\documentclass{book}

\usepackage{graphicx}
\usepackage{graphics}
\usepackage{epsfig}
\usepackage{amsmath}
\usepackage{amsfonts}
\usepackage{amssymb}
\usepackage{tikz}
\font\rs=cmss10.360pk
\font\rt=cmss9.360pk
\font\sd=cmcsc9.360pk

\include {mak}
\begin{document}

%Your \newcommands below (if any):

\oddsidemargin 16.5truemm
\evensidemargin 16.5truemm
\thispagestyle{plain}

\noindent{\rs J. Indones. Math. Soc. (MIHMI)}

\vspace{-0.25cc}

\noindent{\scriptsize Vol. 12, No. 1 (2006), pp.~41--57.}

\vspace{5cc}

\begin{center}
{\Large\bf LINEAR THEORY FOR SINGLE AND DOUBLE FLAP WAVEMAKERS\\%avoid formulae in the title
\rule{0mm}{6mm}\renewcommand{\thefootnote}{}\footnotetext{\hspace{-0.55cm}\scriptsize
{\rm Received 15 March 2005, Accepted 27 July 2005.}%Dates are to be filled by editor.
}}

\vspace{2cc}

{\large\sc W.M. Kusumawinahyu, N. Karjanto and G. Klopman}

\vspace{2cc}

\parbox{24cc}{{\scriptsize{\bf Abstract.}
In this paper, we are concerned with deterministic wave generation
in a hydrodynamic laboratory. A linear wavemaker theory is
developed based on the fully dispersive water wave equations. The
governing field equation is the Laplace equation for potential
flow with several boundary conditions: the dynamic and kinematic
boundary condition at the free surface, the lateral boundary
condition at the wavemaker and the bottom boundary condition. In
this work, we consider both single-flap and double-flap wavemakers.
The velocity potential and surface wave elevation are derived, and
the relation between the propagating wave height and wavemaker
stroke is formulated. This formulation is then used to find how to
operate the wavemaker in an efficient way to generate the desired
propagating waves with minimal disturbances near the wavemaker.
\par}}
\end{center}

%Below is the body of the text. Sections should be numbered and their titles
%should be typed in bold. References in the text should read like e.g. Halmos
%[2], or just [2], rather than Halmos (1974).

\vspace{1.5cc}

\begin{center}
{\bf 1. INTRODUCTION}
\end{center}

\begin{figure}[ptb]
\begin{center}
\includegraphics[width=0.7\textwidth]{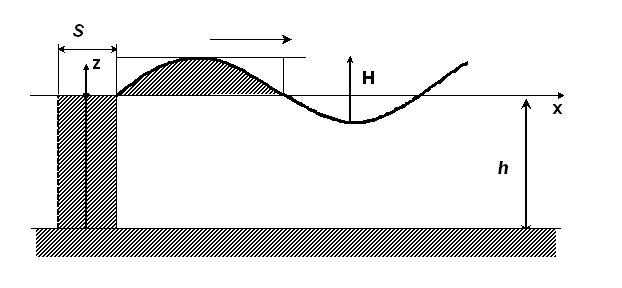}
\caption{Sketch of a two-dimensional wave tank with piston-type of wavemaker.}
\label{Piston}
\end{center}
\end{figure}

\begin{figure}[ptb]
\begin{center}
\includegraphics[width=0.7\textwidth]{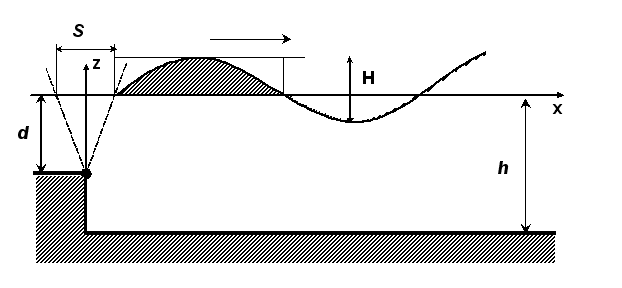}
\caption{Sketch of a two-dimensional wave tank with single-flap
 wavemaker. The hinge is located at $d$ below the still-water
level.} \label{flap}
\end{center}
\end{figure}

A wave tank in a hydrodynamic laboratory is a facility where
maritime structures and ships can be tested by unidirectional
waves on a model scale. It usually has a wavemaker at one side and
a wave-absorbing beach at the other side. Generally, there are two
types of wavemaker which are widely used in hydrodynamic
laboratories, namely piston and flap type wavemakers, as shown in
Figure \ref{Piston} and \ref{flap}. In this paper, we will
consider specifically the flap type of wavemaker which is the
preferred type for testing ships and structures in deep water.
Here deep water means water depth which exceeds approximately a
third of the wavelength. As an example, the Indonesian
Hydrodynamic Laboratory (IHL) in Surabaya, East Java, Indonesia,
uses single and double-flap wavemakers. Flap type wavemakers are
moving partitions which rotate around one or more horizontal axes:
single-flap wavemakers rotate about one hinge elevation, and
double-flap wavemakers have two degrees of freedom. In order to
gain basic insight, we first consider the single flap and then
proceed to the double flap.

For the single-flap wavemaker, suppose that the wave tank has a
still-water depth $h$ and the flap hinge is located at a distance
$d$ below the still-water level. Although the hinge can be located
above or at the bottom or even below the bottom, we will consider
the first case. The bottom of the tank is taken to be flat and
horizontal as well as impermeable. It is assumed that the flap
moves with a monochromatic frequency and can reach a certain
maximum stroke. The generated wave propagates towards the beach,
which is considered to be perfectly absorbing, i.e. no wave is
reflected. The aim is to find out the relation between the stroke
of the wavemaker and the wave height far from the wavemaker. For
this purpose, we need to know the velocity potential and the
surface wave elevation which satisfy the governing equations for
the wave tank. Further, the generation of irregular waves is
considered.

In the case of a double-flap wavemaker, there are two actuators
which move the lower flap (often called the main flap) and the
upper flap. Since we use linear theory, the motion of a
double-flap wavemaker can be regarded as the superposition of two
single flaps. Thus the generated wave is the superposition of the
waves produced by each single flap individually. Once the relation
between the flap motions and the generated waves is determined, we
know how to move the flap in order to generate a certain wave for
testing a ship. The prescribed surface waves can also be irregular
waves having two or many frequencies. Again, linear superposition
can be used to sum solutions with different frequencies. Normally,
the upper flap is used for the high frequencies, while the main
flap is operated to generate the low-frequency waves. Then one
needs to determine the critical frequency to decide for which
frequency range the waves can be generated more efficiently by
either the upper flap or the lower one. A customary way to find
this critical frequency is through the use of the \emph{Figure of
Merit} (FoM), often also called \emph{merit function}. As a
function of frequency $\omega$, the FoM describes the ratio
between the free surface amplitude at the wavemaker, $x = 0$, and
the one at infinity for a monochromatic motion of the wavemaker at
frequency $\omega$.

In the next section we will discuss the single-flap wavemaker
theory. It starts with the simplest theory from Galvin
\cite{Galvin} and then proceeds with the full linear theory for
water waves, with the Laplace equation as governing equation and
the appropriate boundary conditions. Section 3 will explore the
double-flap wavemaker motion, using the theory for a single-flap
obtained in Section 2. The \emph{direct problem}, when the outcome
is the generated wave for a prescribed flap movement, is presented
first. Then we will discuss the \emph{inverse problem}, where the
objective is to determine the motion of the wavemaker in order to
generate the desired surface wave. In that section we also
determine the critical frequency as a criterium to decide whether
to use the main flap or the upper flap. Further, the generation of
irregular waves with a double-flap wavemaker is discussed.
Finally, in Section 4 we will give some conclusions and
recommendations for possible future research on this topic.

\vspace{1.5cc}

\begin{center}
{\bf 2. SINGLE-FLAP WAVEMAKER THEORY}
\end{center}

In this section, we will derive the simple shallow-water wavemaker
theory of Galvin \cite{Galvin}. After that, the linear theory
based on the full equations for water wave motion is discussed in
more detail. By finding the ratio between the wave height and the
wavemaker stroke, we solve the direct problem and the inverse
problem at the same time. With the direct problem we mean that by
prescribing the wavemaker stroke, the wave height far from the
wavemaker follows as an outcome. The inverse problem is that if we
desire to make a certain wave height far away from the wavemaker, we can
calculate the stroke needed as an input to the wavemaker.

\vspace{0.75cc}
\begin{center}
{\bf 2.1 Galvin's theory for linear shallow-water waves}
\end{center}

Consider a flap type wavemaker as shown at Figure \ref{flap}.
Galvin \cite{Galvin} proposed a simple theory for the generation
of waves by this wavemaker, valid in shallow-water. Shallow-water
means wavelengths $L$ much larger than the water depth $h$, say
$L > 10 h$. He reasons that the water displaced by a full stroke
of the wavemaker should be equal to the crest volume of the
propagating wave. For a flap type of wavemaker with a maximum
stroke $S$ at the still-water level and a hinge depth $d$, the
volume of water displaced during a whole stroke is
$\frac{1}{2}Sd$. The volume of water in a wave crest for a wave
with height $H$ and wave number $k = \frac{2\pi}{L}$, is given by
\[ \int_{0}^{L/2} \frac{1}{2}H \sin kx \, dx =
\frac{1}{2}\frac{H}{k}
  \left(1 - \cos \frac{1}{2}k L \right)
  = \frac{H}{k}.\]
By equating the two volumes, we obtain the ratio of wave height
$H$ and the stroke $S$, given by
\begin{equation}
\left(\frac{H}{S}\right)_{\textmd{\tiny Galvin}} = \frac{1}{2} kd.
\label{HSflap}
\end{equation}
However, since our main objective is deep-water waves, Galvin's
theory is not of direct use for our applications. But it can be
used to check the asymptotic behaviour of our results for $kh \ll 1$.

\vspace{0.75cc}
\begin{center}
{\bf 2.2 Linear theory for arbitrary water depth}
\end{center}

Let us start with the mass conservation equation for an
\emph{inviscid} fluid, with mass density $\rho$ and velocity
vector $\boldsymbol{u}$:
%\begin{equation*}
%\nonumber
\[ \frac{\partial \rho}{\partial t} + \nabla (\rho \boldsymbol{u}) = 0.\]
%\end{equation*}
The fluid is assumed to be \textit{incompressible} and $\rho$ is
taken a constant. Then, the mass conservation equation takes the
simple form $\nabla \cdot \boldsymbol{u} = 0$, which is also knows
as the \textit{continuity equation}. Furthermore, to good
approximation for water waves, the motion may be taken to be
\textit{irrotational}, which physically means that individual
fluid particles do not rotate. Mathematically, this implies that
the vorticity vanishes, $\nabla \times \boldsymbol{u} =
\boldsymbol{0}$. Then, there exists a single-valued velocity
potential $\Phi$ such that $\boldsymbol{u} = \nabla \Phi$. By
combining these two assumptions (incompressibility and
irrotationality), we get the \textit{Laplace equation} $\nabla^{2}
\Phi = 0$ as a governing equation for the water wave motion. For
our wavemaker problem, it reads
\[\frac{\partial^{2}\Phi}{\partial x^{2}} + \frac{\partial^{2}\Phi}{\partial z^{2}}=0,
  \hspace{1cm} -h \leq z \leq \eta(x,t), \; x \geq s(z,t).\]
The boundary conditions are as follows
\begin{itemize}
 \item the bottom boundary condition:
 \[\frac{\partial\Phi}{\partial z} = 0, \hspace{1cm} \textmd{at} \; z
 = -h;\]

 \item the free-surface kinematic boundary condition
 \[\frac{\partial\eta}{\partial t} + \frac{\partial\Phi}{\partial
 x}\frac{\partial\eta}{\partial x} = \frac{\partial\Phi}{\partial
 z}, \hspace{1cm} \textmd{at} \; z = \eta(x,t); \]

 \item the free-surface dynamic boundary condition
 \[g\eta + \frac{\partial\Phi}{\partial t} +
 \frac{1}{2}\left[\left(\frac{\partial\Phi}{\partial x}\right)^{2}
 + \left(\frac{\partial\Phi}{\partial z}\right)^{2}\right] = 0,
 \hspace{1cm} \textmd{at} \; z = \eta(x,t); \]

 \item the lateral boundary condition at the wavemaker
 \[\frac{\partial\Phi}{\partial z} \frac{\partial s}{\partial z} + \frac{\partial s}{\partial t} =
 \frac{\partial\Phi}{\partial x} , \hspace{0.2cm} \textmd{at} \; x = s(z,t).\]
\end{itemize}
\begin{figure}[ptb]
\begin{center}
\includegraphics[width=0.7\textwidth]{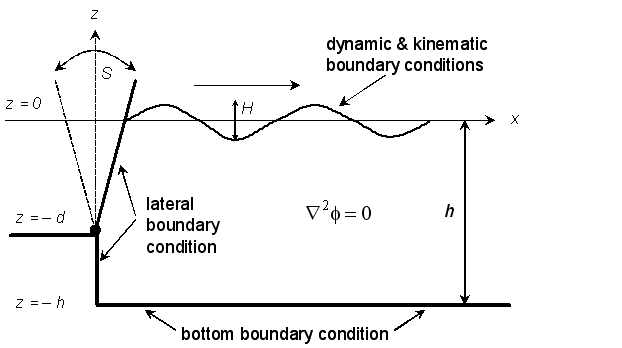}
\caption{The single-flap wavemaker with the governing equation and
its boundary conditions.} \label{flap1}
\end{center}
\end{figure}

Therefore we have a boundary value problem. Figure \ref{flap1}
illustrates the governing equation and its boundary conditions for
a single-flap wavemaker.

Except for the bottom boundary condition, all boundary conditions
are nonlinear. Generally, it is hard to find the exact solution
for this boundary value problem. The solution will be approximated
by linearizing the nonlinear boundary conditions. The linearized
equation model for our boundary value problem now reads
\[\frac{\partial^{2}\Phi}{\partial x^{2}} +
\frac{\partial^{2}\Phi }{\partial z^{2}} = 0,\hspace{1.9cm} -h
\leq z\leq 0,\]
\begin{eqnarray}
\frac{\partial \Phi}{\partial z} &=& 0, \hspace{2.1cm} \hbox{\textmd{at} \; z = -h;} \nonumber \\
\frac{\partial \eta}{\partial t} &=& \frac{\partial\Phi}{\partial z}, \hspace{1.7cm} \hbox{\textmd{at} \; z = 0;} \nonumber \\
\eta + \frac{1}{g}\frac{\partial\Phi}{\partial t} &=& 0,
\hspace{2.1cm} \hbox{\textmd{at} \; z = 0;}
\label{LinDin BC} \\
\frac{\partial\Phi}{\partial x} &=& \frac{\partial
s(z,t)}{\partial t}, \hspace{1.1cm} \hbox{\textmd{at} \; x = 0.}
\label{LinLatBC}
\end{eqnarray}
The lateral boundary motion $s(z,t)$ for a single-flap wavemaker and sinusoidal flap motion becomes:
\[s(z,t) = \frac{1}{2} {\cal S}(z) \sin (\omega t + \psi) = \left\{
 \begin{array}{ll}
 \frac{1}{2} S \left(1 + \frac{z}{d} \right) \sin (\omega t +
 \psi) & \hbox{$, -d \leq z \leq 0;$}\\
 \\
 0 & \hbox{$,-h \leq z \leq -d$,}
 \end{array}
 \right.\] which describes the wavemaker motion with maximum stroke $S$, wavemaker
 frequency $\omega$ and phase $\psi$.

\vspace{0.75cc}
\begin{center}
{\bf 2.3 Velocity potential and wave height-stroke relationship}
\end{center}

\begin{figure}[h]
\begin{center}
\begin{tikzpicture}[scale=0.75]
\draw[->] (0,0) -- (5.5,0) node[anchor=north west] {\footnotesize{$kh$}};
\draw[->] (0,-1.5) -- (0,3.5) node[anchor=east] {\footnotesize{$f_1$}};
\foreach \x in {1,2,3,4,5} \draw (\x cm,1pt) -- (\x cm,-1pt) node[anchor=north] {\footnotesize{$\x$}};
\foreach \y in {-1,0,1,2,3}       \draw (1pt,\y cm) -- (-1pt,\y cm) node[anchor=east] {\footnotesize{$\y$}};
\draw[dashed,domain=0.3:5.2,blue] plot (\x,{1/\x});
\draw[red, domain=0.01:5.2] plot (\x,{tanh(\x)});
\draw node at (5,3.5) {\footnotesize{(a)}};

\begin{scope}[shift={(8,0)}]
\draw[->] (0,0) -- (7.5,0) node[anchor=north west] {\footnotesize{$\kappa h$}};
\draw[->] (0,-1.5) -- (0,3.5) node[anchor=east] {\footnotesize{$f_2$}};
\foreach \x in {1,2,3,4,5,6,7} \draw (\x cm,1pt) -- (\x cm,-1pt) node[anchor=north] {\footnotesize{$\x$}};
\foreach \y in {-1,0,1,2,3}       \draw (1pt,\y cm) -- (-1pt,\y cm) node[anchor=east] {\footnotesize{$\y$}};
\draw[dashed,domain=0.3:7.2,blue] plot (\x,{1/\x});
\draw[red, domain=0.01:1.1] plot (\x,{-tan(deg(\x))});
\draw[red, domain=1.85:4.25] plot (\x,{-tan(deg(\x))});
\draw[red, domain=4.98:7.4] plot (\x,{-tan(deg(\x))});
\draw node at (7,3.5) {\footnotesize{(b)}};
\end{scope}
\end{tikzpicture}
\end{center}
\caption{(a) Graphical representation of the linear dispersion
relation for the propagating modes, the dash-dot curve
represents the plot of $\frac{\sigma^{2}}{kh}$ while the solid one
represents $\tanh kh$.
\newline (b) Graphical representation of the linear dispersion relation
for the evanescent modes, showing four of the infinite numbers of
roots. The dash-dot curve represents the plot of
$\frac{\sigma^{2}}{\kappa h}$, while the solid ones represent
$-\tan \kappa h$.} \label{Propagate_Evanescent}
\end{figure}
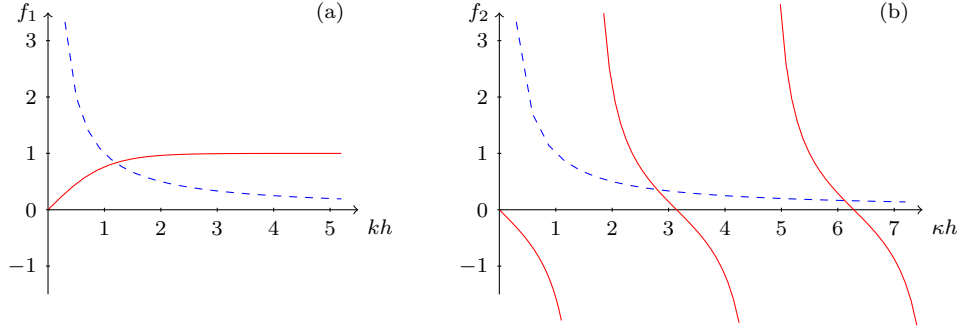

The general solution for the Laplace equation with the bottom and
the free-surface boundary conditions can be determined using the
method of separation of variables; and is given by
\begin{eqnarray}
 \Phi(x,z,t) &=& \frac{g}{\omega}\left[ A \frac{\cosh
k(h+z)}{\cosh kh} \sin(kx-\omega t-\psi) \right] \nonumber\\
 &+& \frac{g}{\omega}\left[ C e^{-\kappa x} \frac{\cos \kappa
(h+z)}{\cos \kappa h} \cos(\omega t + \psi)\right] \nonumber.
\end{eqnarray}
The first term is associated with a \textit{progressive wave}
(also called \textit{propagating mode}), while the second term is
associated with a spatially decaying \textit{standing wave} and is often called an
\textit{evanescent mode}. The wave number $k$ of a progressive wave
and the wave number $\kappa$ of an evanescent mode are related to
the frequency $\omega$ by the \textit{linear dispersion relation}
\begin{equation}
\omega^{2} = gk \tanh kh \label{ProgDisper}
\end{equation}
and
\begin{equation}
\label{EvaDisper}\omega^{2} = -g \kappa \tan \kappa h.
\end{equation}
By rewriting equation (\ref{ProgDisper}) and (\ref{EvaDisper}) as
\[\frac{\sigma^{2}}{kh}=\tanh kh \]
and
\[\frac{\sigma^{2}}{\kappa h}=-\tan \kappa h,\]
where $\sigma = \sqrt{\frac{\omega^2 h}{g}}$ is the
non-dimensional frequency, it is easy to make plots of the
solutions of these equations. The respective plots for $\sigma =
1$ are given in Figure \ref{Propagate_Evanescent}.

Since there is an infinite number of solutions to equation
(\ref{EvaDisper}), the solution for the boundary value
problem has to be written as
\begin{eqnarray}
\Phi(x,z,t) &=& \frac{g}{\omega} \left[A \frac{\cosh
k(h+z)}{\cosh k h} \sin(kx-\omega t-\psi) \right] \nonumber\\
&+& \frac{g}{\omega} \left[ \sum_{n=1}^{\infty} C^{[n]} e^{-\kappa
^{[n]} x} \frac{\cos \kappa ^{[n]} (h+z)}{\cos \kappa ^{[n]} h} \cos(\omega t
+ \psi) \right], \nonumber
\end{eqnarray}
where $A$ and $C^{[n]}$ need to be determined. We assume all wave
motion to originate from the wavemaker, so we only consider the
positive $k>0$ and $\kappa^{[n]} > 0$ to the dispersion relation
(\ref{ProgDisper}) and (\ref{EvaDisper}). Note that the evanescent modes
with amplitudes $C^{[n]}, n = 1, 2, ...$ decay to zero far away
from the wavemaker. Substitute this velocity potential into the
linearized lateral boundary condition (\ref{LinLatBC}) to get
\begin{eqnarray}
& & \hspace*{-0.9cm} \frac{g}{\omega} \left[\frac{Ak}{\cosh kh} \cosh k(h + z) -
\sum_{n=1}^{\infty} \frac{C^{[n]} \kappa ^{[n]}}{\cos \kappa ^{[n]}
h } \cos \kappa ^{[n]} (h + z) \right] \nonumber \\
&=& \left\{
\begin{array}{ll}
\frac{1}{2}S \omega \left(1 + \frac{z}{d} \right)  & \hbox{$,-d \leq z \leq 0;$}\\
\\
0 & \hbox{$,-h \leq z \leq -d,$}
\end{array}
\right. \label{SubsLatBC}
\end{eqnarray}
which has to be valid for any $-h \leq z \leq 0.$

It is known from the Sturm-Liouville condition that the set
\[\left\{\cosh k(h+z),\cos \kappa^{[n]} (h+z), n=1,2,\dots\right\}\]
forms an orthogonal set, namely
\[ \int_{-h}^{0}\cosh k(h+z) \cos \kappa ^{[n]} (h+z) dz=0 \]
and
\[\int_{-h}^{0}\cos \kappa ^{[n]} (h+z) \cos \kappa^{[m]} (h+z) dz=0, \hspace{0.3cm} m\neq
n,\] can be shown to hold because of the linear dispersion relationship
(\ref{ProgDisper}) and (\ref{EvaDisper}).

Therefore, to find $A$ both sides of (\ref{SubsLatBC}) are
multiplied by $\cosh k(h+z)$ and then integrated over the depth.
Due to the orthogonality property, the evanescent terms containing
$C^{[n]}$ are eliminated, and thus
\begin{eqnarray}
A &=& \frac{\sinh kh}{2}\frac{\int_{-d}^{0}\frac{1}{2} \left(1 +
\frac{z}{d} \right) \cosh k(h +
z)dz}{\int_{-h}^{0}\cosh^{2}(k(h+z))dz}S \nonumber \\
&=& 2\left(\frac{\sinh kh}{kd}\right) \frac{\cosh k(h-d) + kd
\sinh kh - \cosh kh}{2kh + \sinh 2kh} S.
\label{Ap}
\end{eqnarray}

To find $C^{[n]}$, multiply equation (\ref{SubsLatBC}) by $\cos
\kappa ^{[n]} (h+z)$ and then integrate over the depth. It is found
that
\begin{eqnarray}
C^{[n]} &=& \frac{\sin \kappa ^{[n]} h}{2} \frac{\int_{-h}^{0}
\frac{1}{2}\left(1 + \frac{z}{d} \right) \cos \kappa ^{[n]} (h+z) dz}
{\int_{-h}^{0}\cos^{2}(\kappa ^{[n]}
(h+z))dz} S \nonumber \\
&=& -2 \left(\frac{\sin \kappa ^{[n]} h}{\kappa ^{[n]} d}\right)
\frac{\cos \kappa ^{[n]}(h-d) - \kappa ^{[n]} d \sin \kappa ^{[n]} h - \cos
\kappa ^{[n]} h}{2 \kappa ^{[n]} h + \sin 2 \kappa ^{[n]} h} S.
\label{Cn}
\end{eqnarray}

Finally, the elevation of the generated surface wave $\eta(x,t)$
is readily found by applying the linearized dynamic boundary
condition at the still-water level, equation (\ref{LinDin BC}),
namely
\begin{eqnarray}
\eta = -\frac{1}{g}\frac{\partial\Phi}{\partial t} \Bigg|_{z=0}
&=& A \cos(kx - \omega t-\psi) + \sum_{n=1}^{\infty} C^{[n]}
e^{-\kappa ^{[n]} x} \sin (\omega t + \psi) \nonumber \\
& =& \frac{H}{2} \cos (kx-\omega t-\psi)+ \sum_{n=1}^{\infty}
C^{[n]} e^{-\kappa ^{[n]} x} \sin(\omega t + \psi), \nonumber
\label{etaSingle}
\end{eqnarray}
so $A$ is the progressive amplitude and $H = 2A$ is the wave height.
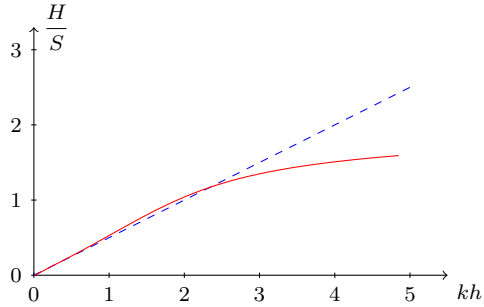
\begin{figure}[h]
\begin{center}
\begin{tikzpicture}[scale=1]
\draw[->] (0,0) -- (5.5,0) node[anchor=north west] {\footnotesize{$kh$}};
\draw[->] (0,0) -- (0,3.3) node[anchor=west] {\footnotesize{${\displaystyle \frac{H}{S}}$}};
\foreach \x in {0,1,2,3,4,5} \draw (\x cm,1pt) -- (\x cm,-1pt) node[anchor=north] {\footnotesize{$\x$}};
\foreach \y in {0,1,2,3}       \draw (1pt,\y cm) -- (-1pt,\y cm) node[anchor=east] {\footnotesize{$\y$}};
\draw[dashed,domain=0:5,blue] plot (\x,{\x/2});
\draw[red, domain=0.01:4.85] plot (\x,{4*sinh(\x)/(\x)*(1 + (\x)*sinh(\x) - cosh(\x))/(2*(\x) + sinh(2*(\x)))});
\end{tikzpicture}
\end{center}
\caption{Wave height to stroke ratio $H/S$ as a function
of $kh$ for a flap type wavemaker, with $d = h$, based on Galvin's theory,
equation (\ref{HSflap}), (dash-dot curve), and the full linear theory, equation
(\ref{TransfFlap1}), (solid curve).} \label{ratioHS}
\end{figure}

The wave height of the progressive wave is determined by
evaluating $\eta(x,t)$ far from wavemaker, where the evanescent
modes vanish, namely
\[ \eta = \frac{H}{2}\cos(kx-\omega t-\psi), \hspace{0.25cm} \kappa^{[1]} x \gg 1.\]

Since the relation between $A$ and $S$ is known, the ratio of
the wave height $H = 2A$ and the stroke $S$ is given by
\begin{equation}
\left(\frac{H}{S}\right)_{\textmd{\tiny linear}} =
4\left(\frac{\sinh kh}{kd}\right) \frac{\cosh k(h-d) + kd \sinh kh
- \cosh kh}{2kh + \sinh 2kh}.
\label{TransfFlap}
\end{equation}
When the hinge of the flap is located at the bottom of the wave tank
we have $d=h$ and the ratio becomes
\begin{equation}
\left(\frac{H}{S}\right)_{\textmd{\tiny linear}} =
4\left(\frac{\sinh kh}{kh}\right) \frac{1 + kh \sinh kh - \cosh
kh}{2kh + \sinh 2kh},
\label{TransfFlap1}
\end{equation}
as also found in Dean and Dalrymple \cite{Dean} and Gilbert et al
\cite{Gilbert}. Figure \ref{ratioHS} shows the plots of wave
height to stroke ratio for flap type of wavemaker using the full
linear theory, equation (\ref{TransfFlap1}), compared to Galvin's
theory, equation (\ref{HSflap}), when $d=h$. Therefore Galvin's
simple theory gives a good approximation to the wave height to
stroke ratio for small $kh$.

The wave height to stroke ratio (\ref{TransfFlap}) gives the
transformation from the stroke of the single-flap wavemaker motion
to the progressive wave height generated by that wavemaker motion.
For the testing of ships in a hydrodynamic laboratory, normally
the wave height $H$ of a monochromatic signal is specified. Using
(\ref{TransfFlap}), we now can determine the required stroke of
the wavemaker.

Often, the desired waves are not regular but irregular. These
waves can be represented as a summation of several monochromatic
waves, namely
\begin{equation}
\eta(x,t) = \sum_{i=1}^{N} \frac{1}{2} H_{i} \cos(k_i x -
\omega_{i}t - \psi_{i}), \label{DesiredEta}
\end{equation}
where each wave number - frequency pair $(k_i, \omega_i)$ satisfies the
linear dispersion relation (\ref{ProgDisper}). Since we consider
linear wavemaker theory the wavemaker has to be moved with the
flap motion
\[s(z,t)=\left\{
\begin{array}{ll}
{\displaystyle \frac{1}{2} \left(1 + \frac{z}{d}\right)
\sum_{i=1}^{N} S_{i}\sin(\omega_{i}t + \psi_{i} )} & \hbox{$,-d \leq z \leq  0;$}\\
0 & \hbox{$, -h \leq z \leq -d,$}
\end{array}
\right.\] where every $S_{i}$ satisfies the wave height to stroke
ratio (\ref{TransfFlap}) associated with the corresponding wave height $H_{i}$.

\vspace{1.5cc}
\begin{center}
{\bf 3. DOUBLE-FLAP WAVEMAKER THEORY}
\end{center}

This section is started by the formulation of the double-flap
wavemaker motion and the surface waves generated by the flap
movement. This formulation is referred to as the direct problem.
Once the relation between the flap strokes and the height of the
generated progressive wave is found we proceed with the inverse
problem. There, the desired surface wave at a position far from
the wavemaker is given, generally in the form of an irregular
wave. Hence it can be represented as a summation of a finite
number of regular waves with different frequencies and amplitudes.
Regular waves with high frequencies are generated by moving the
upper flap, while the low-frequency waves are generated by the
main flap. Thus we need to find the critical frequency that
separates the frequency range for which one shall operate either
the upper flap or the main flap.

\vspace{0.75cc}
\begin{center}
{\bf 3.1 The generated surface waves}
\end{center}

Consider a wave tank with a double-flap wavemaker at one side and
a wave-absorbing beach at the other side, as shown at Figure
\ref{Schema_doubleflap}. Suppose that the depth of the tank is $h$
and the hinges of the upper flap and main flap are located at
distances $d_{1}$ and $d_{2}$ respectively below the still-water
level. If the upper flap moves with frequency $\omega_{1}$ and has
a maximum stroke $S_{1}$ and the main flap moves with frequency
$\omega_{2}$ and has a maximum stroke $S_{2}$, then we can
distinguish three cases for the motion of the wavemaker: upper
flap motion, main flap motion, and combined flap motion. Figure
\ref{3cases} illustrates these three situations. Since we consider
only the linear theory of wavemakers, then the last case is just
the superposition of the two preceding cases.
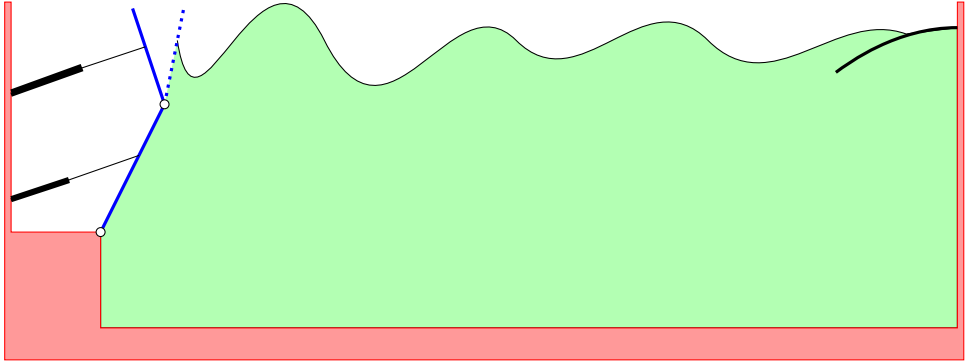
\begin{figure}[h]
\begin{center}
\begin{tikzpicture}[scale=0.85]
\filldraw[fill = green!30!white] (2.7,5) .. controls (3,3) and (4,7) .. (5,5) .. controls (6,3) and (7,6) .. (8,5) .. controls (9,4) and (10,6) .. (11,5) 
.. controls (12,4) and (13,5.5) .. (14.1,5.1) -- (14.9,5.2) -- (14.9,0.5) -- (1.5,0.5) -- (1.5,2) -- (2.5,4);
\filldraw[fill=red!40!white, draw=red] (0,0) -- (15,0) -- (15,5.6) -- (14.9,5.6) -- (14.9,0.5) -- (1.5,0.5) -- (1.5,2) -- (0.1,2) -- (0.1,5.6) -- (0,5.6) -- cycle;
\draw[blue,very thick] (1.5,2) -- (2.5,4) -- (2,5.5);
\draw[blue,very thick, dotted] (2.5,4) -- (2.8,5.5);
\filldraw[fill=white, draw=black] (1.5,2) circle (0.7mm);
\filldraw[fill=white, draw=black] (2.5,4) circle (0.7mm);
\draw[black] (0.1,2.5) -- (2.1,3.2);
\draw[black, ultra thick] (0.1,2.5) -- (1.01,2.8);
\draw[black, ultra thick] (0.1,2.53) -- (1,2.83);
\draw[black] (0.1,4.2) -- (2.2,4.9);
\draw[black, ultra thick] (0.1,4.2) -- (1.2,4.6);
\draw[black, ultra thick] (0.1,4.15) -- (1.22,4.55);
%\draw (2.7,5) .. controls (3,3) and (4,7) .. (5,5) .. controls (6,3) and (7,6) .. (8,5) .. controls (9,4) and (10,6) .. (11,5) 
%.. controls (12,4) and (13,5.5) .. (14.1,5.1);
\draw[black, very thick] (14.9,5.2) parabola (13,4.5);
\end{tikzpicture}
\end{center}
\caption{Schematic figure of a double-flap wavemaker structure.} \label{Schema_doubleflap}
\end{figure}
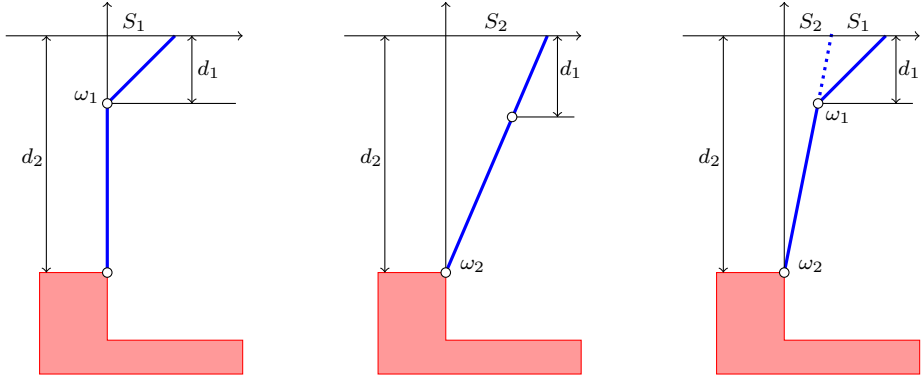
\begin{figure}[h]
\begin{center}
\begin{tikzpicture}[scale=0.9]
\filldraw[fill=red!40!white, draw=red] (0,0) -- (3,0) -- (3,0.5) -- (1,0.5) -- (1,1.5) -- (0,1.5) -- cycle;
\draw[blue,very thick] (1,1.5) -- (1,4) -- (2,5);
\draw[->] (1,4) -- (1,5.5);
\draw node[black] at (1.4,5.2) {\footnotesize{$S_1$}};
\draw (1,4) -- (2.9,4);
\filldraw[fill=white, draw=black] (1,1.5) circle (0.7mm);
\filldraw[fill=white, draw=black] (1,4) circle (0.7mm);
\draw node [black] at (0.7,4.1) {\footnotesize{$\omega_1$}};
\draw[->] (-0.5,5) -- (3,5);
\draw[<->] (0.1,1.5) -- (0.1,5);
\draw node [black] at (-0.1,3.2) {\footnotesize{$d_2$}};
\draw[<->] (2.25,4) -- (2.25,5);
\draw node [black] at (2.5,4.5) {\footnotesize{$d_1$}}; 

\begin{scope}[shift={(5,0)}]
\filldraw[fill=red!40!white, draw=red] (0,0) -- (3,0) -- (3,0.5) -- (1,0.5) -- (1,1.5) -- (0,1.5) -- cycle;
\draw[blue,very thick] (1,1.5) -- (2.5,5);
\draw[->] (1,1.5) -- (1,5.5);
\draw (2,3.8) -- (2.9,3.8);
\filldraw[fill=white, draw=black] (1,1.5) circle (0.7mm);
\filldraw[fill=white, draw=black] (1.98,3.8) circle (0.7mm);
\draw node [black] at (1.4,1.6) {\footnotesize{$\omega_2$}};
\draw[->] (-0.5,5) -- (3,5);
\draw node[black] at (1.75,5.2) {\footnotesize{$S_2$}};
\draw[<->] (0.1,1.5) -- (0.1,5);
\draw node [black] at (-0.1,3.2) {\footnotesize{$d_2$}}; 
\draw[<->] (2.65,3.8) -- (2.65,5);
\draw node [black] at (2.85,4.5) {\footnotesize{$d_1$}}; 
\end{scope}

\begin{scope}[shift={(10,0)}]
\filldraw[fill=red!40!white, draw=red] (0,0) -- (3,0) -- (3,0.5) -- (1,0.5) -- (1,1.5) -- (0,1.5) -- cycle;
\draw[blue,very thick, dotted] (1.5,4) -- (1.7,5.03);
\draw[blue,very thick] (1,1.5) -- (1.5,4) -- (2.5,5);
\draw[->] (1,1.5) -- (1,5.5);
\draw (1.5,4) -- (2.9,4);
\filldraw[fill=white, draw=black] (1,1.5) circle (0.7mm);
\filldraw[fill=white, draw=black] (1.5,4) circle (0.7mm);
\draw node [black] at (1.8,3.8) {\footnotesize{$\omega_1$}};
\draw node [black] at (1.4,1.6) {\footnotesize{$\omega_2$}};
\draw[->] (-0.5,5) -- (3,5);
\draw node[black] at (1.4,5.2) {\footnotesize{$S_2$}};
\draw node[black] at (2.1,5.2) {\footnotesize{$S_1$}};
\draw[<->] (0.1,1.5) -- (0.1,5);
\draw node [black] at (-0.1,3.2) {\footnotesize{$d_2$}}; 
\draw[<->] (2.65,4) -- (2.65,5);
\draw node [black] at (2.85,4.5) {\footnotesize{$d_1$}}; 
\end{scope}
\end{tikzpicture}
\end{center}
\caption{\small{Illustration of three cases in the double flaps
wavemaker: upper flap motion (left), main flap motion (middle),
combination of upper and main flap (right).}} \label{3cases}
\end{figure}

The cases when only the upper flap moves with stroke $S_1$ or the
main flap with stroke $S_2$ has been treated in the previous
section on single-flap motion. Here, we will discuss the case of
both flaps moving simultaneusly, the upper flap with frequency
$\omega_1$ and the main flap with frequency $\omega_2$.

The horizontal displacement $s(z,t)$ when both the upper flap and
the main flap move together is given by
\[ s(z,t) = \left\{%
\begin{array}{ll}
    \frac{1}{2}\left[S_{1}\left(1 + \frac{z}{d_{1}}\right) \sin(\omega_{1}t + \psi_{1}) +
    S_{2} \left(1+\frac{z}{d_{2}}\right) \sin(\omega_{2}t + \psi_{2})\right] & \hbox{$,-d_{1} \leq z \leq 0;$} \\
    \\
    \frac{1}{2} S_{2} \left(1+\frac{z}{d_{2}}\right) \sin(\omega_{2}t + \psi_{2}) & \hbox{$,-d_{2} \leq z \leq -d_{1};$} \\
    \\
    0 & \hbox{$,-h \leq z \leq -d_{2}.$}
\end{array}%
\right. \]

By linear superposition, the corresponding velocity potential
$\Phi$ for this condition is the sum of the velocity potentials
$\Phi_{1}$ and $\Phi_{2}$, namely
\begin{eqnarray}
  \Phi(x,z,t) &=& \frac{g}{\omega_{1}} \left[A_{1} \frac{\cosh k_{1}(z + h)}{\cosh k_{1}h} \sin (k_{1}x - \omega_{1}t - \psi_{1}) \right] \nonumber \\
  &+& \frac{g}{\omega_{1}} \left[\sum_{n = 1}^{\infty} C_{1}^{[n]} e^{-\kappa_{1}^{[n]}x} \frac{\cos \kappa_{1}^{[n]}(z + h)}{\cos \kappa_{1}^{[n]}h} \cos(\omega_{1}t + \psi_{1}) \right] \nonumber \\
  &+& \frac{g}{\omega_{2}} \left[A_{2} \frac{\cosh k_{2}(z + h)}{\cosh k_{2}h} \sin (k_{2}x - \omega_{2}t - \psi_{2})\right] \nonumber \\
  &+& \frac{g}{\omega_{2}} \left[\sum_{n = 1}^{\infty} C_{2}^{[n]} e^{-\kappa_{2}^{[n]}x} \frac{\cos \kappa_{2}^{[n]}(z + h)}{\cos \kappa_{2}^{[n]}h} \cos(\omega_{2}t + \psi_{2})
  \right], \nonumber
\end{eqnarray}
where the coefficients $A_{1}$, $A_{2}$, $C_{1}^{[n]}$, and
$C_{2}^{[n]}$ are as given for the single-flap case. The surface
elevation $\eta(x,t)$ produced by the double-flap wavemaker
becomes, according to the linearized dynamic boundary condition
(\ref{LinDin BC}):
\begin{eqnarray}
  \eta(x,t) &=& -\frac{1}{g} \frac{\partial \Phi}{\partial t} \bigg|_{z = 0} \nonumber \\
  &=& A_{1} \cos (k_{1}x - \omega_{1}t - \psi_{1}) + \sum_{n = 1}^{\infty} C_{1}^{[n]} e^{-\kappa_{1}^{[n]}x} \sin (\omega_{1}t + \psi_{1}) \nonumber \\
  &+& A_{2} \cos (k_{2}x - \omega_{2}t - \psi_{2}) + \sum_{n = 1}^{\infty} C_{2}^{[n]} e^{-\kappa_{2}^{[n]}x} \sin (\omega_{2}t +
  \psi_{2}).\nonumber
\end{eqnarray}

The generated progressive wave is determined by evaluating
$\eta(x,t)$ far from wavemaker, where the evanescent modes vanish:

\begin{eqnarray}
  \eta(x,t) &=& A_{1} \cos (k_{1}x - \omega_{1}t - \psi_{1}) +  A_{2} \cos (k_{2}x - \omega_{2}t - \psi_{2}) \nonumber \\
            &=& \frac{H_{1}}{2} \cos (k_{1}x - \omega_{1}t - \psi_{1}) +  \frac{H_{2}}{2} \cos (k_{2}x - \omega_{2}t -
            \psi_{2}), \nonumber
\end{eqnarray}
where
\[ H_{1} = 4 \left( \frac{ \sinh k_{1}h}{k_{1}d_{1}} \right) \frac{\cosh k_{1}(h-d_{1}) + k_{1}d_{1} \sinh k_{1}h - \cosh
k_{1}h}{2k_{1}h + \sinh 2k_{1}h} S_{1}, \] and
\[ H_{2} = 4 \left(\frac{\sinh k_{2}h}{k_{2} d_{2}} \right) \frac{\cosh k_{2}(h-d_{2})
      + k_{2}d_{2} \sinh k_{2}h - \cosh k_{2}h}{2k_{2}h + \sinh 2k_{2}h}
      S_2, \]
relate the wave heights $H_1$ and $H_2$ to the strokes $S_1$ and
$S_2$, respectively. Note, that for the case $\omega_1 \neq
\omega_2$, the maximum wave height is $H_1 + H_2$ and the minimum
wave height is $|H_1 - H_2|$ in the generated bi-chromatic wave
pattern.

\vspace{0.75cc}
\begin{center}
{\bf 3.2 Inverse Problem}
\end{center}

In the previous subsection, the formulations give the heights of
the generated waves for a prescribed flap motion. The wave heights
are expressed in terms of the wavemaker stroke, the hinge
positions, water depth and the frequencies of the flap motion. In
this subsection we want to determine how to move the flaps in such
a way such that the desired waves are generated. Normally, the
upper flap generates the high-frequency waves, while the main flap
generates the low-frequency ones. Generally, the desired waves are
not regular waves and are written as a linear superposition of
many regular waves with various frequencies. Hence one needs to
find the critical frequency above which one uses the upper flap
and below which one uses the main flap to generate waves. For this
reason, the next subsection will discuss a method to determine the
critical frequency.

\vspace{0.75cc}
\begin{center}
{\bf 3.2.1 The critical frequency}
\end{center}

A \textit{merit} function, also known as a
\textit{Figure-of-Merit} function, is a function that measures the
agreement between data and a model fitted with for a particular choice
of the parameters. By convention, the merit function is small when
the agreement is good. In the process known as regression,
parameters are adjusted based on the value of the merit function
until a minimum is obtained, thus producing a best-fit. The
corresponding parameters, giving the smallest value of the merit
function, are known as the best-fit parameters (Press
\cite{Press}).

Our aim is to determine for which frequency range the upper flap
and the main flap will be used. Therefore, we need a method to
solve this problem. We introduce the Figure of Merit (FoM) of a
wavemaker flap, which is given by the ratio of the surface wave
amplitude at the wavemaker and the one far from the wavemaker. It
reads (Dalzell \cite{Dalzell}):
\[\textmd{FoM}(\omega) = \frac{\sqrt{A^{2}(\omega) + \left[\sum_{n = 1}^{\infty} C^{[n]}(\omega)
  \right]^{2}}}{A(\omega)},\]
where $A$ is the propagating mode amplitude and $C^{[n]}$ are the
evanescent mode amplitudes. Note that the FoM is always larger or
equal to one. The ideal case is if the FoM equals one. Then there
are no evanescent modes, which is desirable since the increased
wave height near the wavemaker, due to the presence of the
evanescent modes, may for instance trigger undesirable wave
breaking near the wavemaker. Let the FoM of the upper flap be
denoted by FoM$_{1}$ and the FoM of the main flap be denoted by
FoM$_{2}$, then there exists a critical frequency $\omega^{\ast}$
such that FoM$_{1}(\omega^{\ast})$ = FoM$_{2}(\omega^{\ast})$. The
plots of FoM$_{1}(\omega^{\ast})$ and FoM$_{2}(\omega^{\ast})$ can
easily be drawn when we have found the progressive wavenumber $k$
and evanescent wave numbers $\kappa^{[n]}$ corresponding to a
certain range of $\omega$ by solving the \textit{linear dispersion
relation} (\ref{ProgDisper}) and (\ref{EvaDisper}).

For low frequencies the upper flap FoM$_{1}(\omega)$ will be
higher than the main flap FoM$_{2}(\omega)$, and above a critical
frequency $\omega^{\ast}$, it will be just opposite. The critical
frequency is obtained where the plotted lines of the FoM cross
each other. Accordingly, for $0 < \omega \leq \omega^{\ast}$, the
main flap is used, and for $\omega > \omega^{\ast}$, the upper flap is used.

The following derivation shows that the determination of the
critical frequency depends only on the configuration of wavemaker,
namely the wave tank depth $h$ and the location of the hinges
$d_1$ as well as $d_2$ below the still-water level. Starting from
FoM$_{1}(\omega^{\ast})$ = FoM$_{2}(\omega^{\ast})$ and taking the
square of both sides, we obtain
\begin{eqnarray}
    \frac{A_{1}^{2}(\omega^{\ast}) + \left[\sum_{n = 1}^{\infty} C_{1}^{[n]}(\omega^{\ast}) \right]^{2}}{A_{1}^{2}(\omega^{\ast})} &=&
    \frac{A_{2}^{2}(\omega^{\ast}) + \left[\sum_{n = 1}^{\infty} C_{2}^{[n]}(\omega^{\ast}) \right]^{2}}{A_{2}^{2}(\omega^{\ast})},
    \; \; \textmd{or} \nonumber\\
    1 + \left[\frac{\sum_{n = 1}^{\infty} C_{1}^{[n]}(\omega^{\ast})}{A_{1}(\omega^{\ast})}\right]^{2} &=&
    1 + \left[\frac{\sum_{n = 1}^{\infty}
    C_{2}^{[n]}(\omega^{\ast}) \nonumber
    }{A_{2}(\omega^{\ast})}\right]^{2} \label{FoM1}.
\end{eqnarray}
By re-arranging the terms and taking the square root of both
sides, we get
\begin{equation}
  \frac{A_{1}(\omega^{\ast})}{A_{2}(\omega^{\ast})} = \pm \frac{\sum_{n = 1}^{\infty} C_{1}^{[n]}(\omega^{\ast})}{\sum_{n = 1}^{\infty}
  C_{2}^{[n]}(\omega^{\ast})}, \label{FoM_wm}
\end{equation}
where $A_{1}(\omega^{\ast}),A_{2}(\omega^{\ast}),
C_{1}^{[n]}(\omega^{\ast})$, and $C_{2}^{[n]}(\omega^{\ast})$ are
related to $S_1$ and $S_2$ through (\ref{Ap}) and (\ref{Cn}):
\begin{eqnarray}
  A_{1}(\omega^{\ast}) &=& 2 \left(\frac{S_{1} \sinh k^{\ast}h}{k^{\ast}
  d_{1}}\right) \frac{\cosh k^{\ast}(h - d_{1}) + k^{\ast}d_{1} \sinh (k^{\ast} h) - \cosh(k^{\ast} h)}{2k^{\ast} h + \sinh (2k^{\ast} h)}, \nonumber \\
  A_{2}(\omega^{\ast}) &=& 2 \left(\frac{S_{2} \sinh k^{\ast}h }{k^{\ast}
  d_{2}} \right) \frac{\cosh k^{\ast}(h - d_{2}) + k^{\ast}d_{2} \sinh (k^{\ast} h) - \cosh(k^{\ast} h)}{2k^{\ast} h + \sinh (2k^{\ast} h)}, \nonumber \\
  C_{1}^{[n]}(\omega^{\ast}) &=& -2 \left(\frac{S_{1} \sin \kappa^{[n]}_{\ast}h}{\kappa^{[n]}_{\ast} d_{1}}
  \right) \frac{\cos \kappa^{[n]}_{\ast}(h - d_{1}) - \kappa^{[n]}_{\ast}d_{1} \sin (\kappa^{[n]}_{\ast} h) - \cos(\kappa^{[n]}_{\ast} h)}
  {2 \kappa^{[n]}_{\ast} h + \sin (2 \kappa^{[n]}_{\ast} h)}, \nonumber \\
  C_{2}^{[n]}(\omega^{\ast}) &=& -2 \left( \frac{S_{2} \sin \kappa^{[n]}_{\ast}h}{\kappa^{[n]}_{\ast} d_{2}}
  \right) \frac{\cos \kappa^{[n]}_{\ast}(h - d_{2}) - \kappa^{[n]}_{\ast}d_{2} \sin (\kappa^{[n]}_{\ast} h) - \cos(\kappa^{[n]}_{\ast} h)}
  {2 \kappa^{[n]}_{\ast} h + \sin (2 \kappa^{[n]}_{\ast} h)}. \nonumber
\end{eqnarray}
By substituting these quantities into (\ref{FoM_wm}), we have the
following relation at $\omega = \omega^{\ast}$:
\begin{eqnarray}
  \frac{\cosh k^{\ast}(h - d_{1}) + k^{\ast} d_{1} \sinh(k^{\ast} h) - \cosh(k^{\ast} h)}
       {\cosh k^{\ast}(h - d_{2}) + k^{\ast} d_{2} \sinh(k^{\ast} h) - \cosh(k^{\ast} h)} =
  \hspace*{6cm} \nonumber \\ \hspace*{-6cm} \pm
  \frac{\sum_{n = 1}^{\infty} \left[\cos \kappa^{[n]}_{\ast}(h - d_{1}) - \kappa^{[n]}_{\ast} d_{1} \sin(\kappa^{[n]}_{\ast} h) - \cos(\kappa^{[n]}_{\ast} h)\right]
                              /[\kappa^{[n]}_{\ast} (2 \kappa^{[n]}_{\ast} h + \sin (2 \kappa^{[n]}_{\ast} h))]}
       {\sum_{n = 1}^{\infty} \left[\cos \kappa^{[n]}_{\ast}(h - d_{2}) - \kappa^{[n]}_{\ast} d_{2} \sin(\kappa^{[n]}_{\ast} h) - \cos(\kappa^{[n]}_{\ast} h)\right]
                             /[\kappa^{[n]}_{\ast} (2 \kappa^{[n]}_{\ast} h + \sin (2 \kappa^{[n]}_{\ast} h))]},
  \label{critical_quantities}
\end{eqnarray}
where $k^{\ast}$, $\kappa^{[n]}_{\ast}$,  and $\omega^{\ast}$
satisfy the linear dispersion relation
\[(\omega^{\ast})^2=g k^{\ast} \tanh(k^{\ast} h) = -
g\kappa^{[n]}_{\ast} \tan(\kappa^{[n]}_{\ast} h).\]
\begin{figure}[ptb]
\begin{center}
\includegraphics[width=0.9\textwidth]{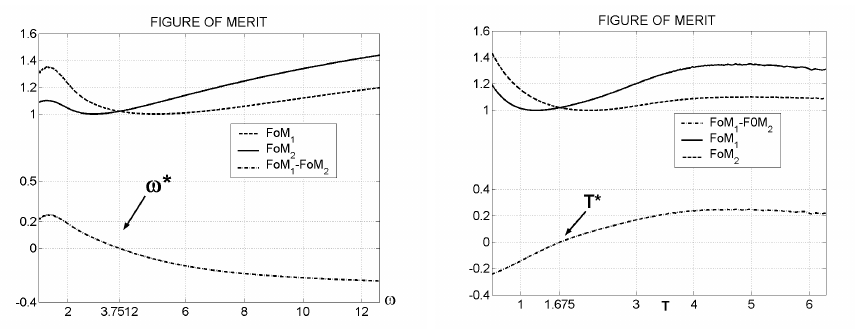}
\caption{\small{Graphs of the Figure of Merit as a function of
frequency (left) and as a function of wave period (right), for $h
= 5.5$ m, $d_1 = 0.83$ m, and $d_2 = 2.55$ m.}} \label{FOM}
\end{center}
\end{figure}

The configuration of the double-flap wavemaker in the towing tank
of the IHL is: $h = 5.5$ m depth and the position of hinges is at
$d_1 = 0.83$ m and $d_2 = 2.55$ m below the still-water level. For
this configuration, we have drawn the plots of FoM$_{1}$ and
FoM$_{2}$ in the frequency domain as well as in the wave period
domain, as shown in Figure \ref{FOM}. It can be noticed that for
the IHL towing tank configuration the critical frequency is
$\omega^{\ast} = 3.7512 $ rad/s, and the critical period is
$T^{\ast} = 2 \pi / \omega^{\ast} = 1.675$ sec. It can also be
observed that at the critical point, the value of FoM$_{1}$ as
well as FoM$_{2}$ are both very close to $1$, the optimum value.
Further it can be seen that if the main flap and upper flap are
used in the appropriate frequency ranges, the FoM is less than
$1.1$ for $T > 0.7$ s $ (\textmd{or} \; \omega < 9$ rad/s).

\newpage
\begin{center}
{\bf 3.2.2 Irregular wave motion for the double-flap wavemaker}
\end{center}

Let the desired irregular waves, to be generated by the
double-flap wavemaker, be expressed by (\ref{DesiredEta}). In
order to distinguish the wave components that have to be generated
by the upper flap from the ones that will be generated by the main
flap, the desired wave field needs to be separated into two parts:
\begin{eqnarray}
\eta(x,t) &=& \sum_{i=1}^{N_1} \frac{H_{1i}}{2}\cos (k_{1i} x -
\omega_{1i} - \psi_{1i}) + \sum_{i=1}^{N_2} \frac{H_{2i}}{2}\cos
(k_{2i} x - \omega_{2i} - \psi_{2i}) \nonumber \\
&=& \eta_{H} (x,t) + \eta_{L} (x,t), \label{SeparatedEta}
\end{eqnarray}
where $\omega_{1i} > \omega^{\ast}$ denote the frequencies of the
wave components that need to be generated by the upper flap, while
$0 < \omega_{2i} \leq \omega^{\ast}$ are the frequencies of the
wave components which will be sent to the main flap. Subscripts
$H$ and $L$ indicate high and low frequency, respectively.

Since we are concerned with linear theory, the flap motion is the
superposition of the flap motion by the individual wave
components. The motion of the double-flap wavemaker to generate
the irregular wave field (\ref{SeparatedEta}) becomes
\[x = \left\{%
\begin{array}{ll}
    \frac{1}{2} \left(1 + \frac{z}{d_{1}}\right) \sum_{i=1}^{N_1} S_{1i}\sin(\omega_{1i}t + \psi_{1i}) \\
     \hspace{2cm} + \frac{1}{2} \left(1 + \frac{z}{d_{2}}\right) \sum_{i=1}^{N_2} S_{2i} \sin(\omega_{2i}t + \psi_{2i}) & \hbox{$,-d_{1} \leq z \leq 0;$} \\
    \\
    \frac{1}{2} \left(1 + \frac{z}{d_{2}}\right) \sum_{i=1}^{N_2} S_{2i} \sin(\omega_{2i}t + \psi_{2i}) & \hbox{$,-d_{2} \leq z \leq -d_{1};$} \\
    \\
    0 & \hbox{$,-h \leq z \leq -d_{2},$}
\end{array}%
\right.\]
where
\[ S_{1i} = \left( \frac{k_{1i}d_{1}}{4 \sinh k_{1i}h } \right) \frac{2k_{1i}h + \sinh 2k_{1i}h}{\cosh k_{1i}(h-d_{1}) + k_{1i}d_{1} \sinh k_{1i}h - \cosh
k_{1i}h} H_{1i}, \]
\[ S_{2i} = \left(\frac{k_{2i} d_{2}}{4 \sinh k_{2i}h } \right) \frac{2k_{2i}h + \sinh 2k_{2i}h}{\cosh k_{2i}(h-d_{2})
      + k_{2i}d_{2} \sinh k_{2i}h - \cosh k_{2i}h} H_{2i},\]
and $k_{ji}$ satisfy the linear dispersion relation
(\ref{ProgDisper}) with respect to $\omega_{ji}$.

\vspace{1.5cc}

\begin{center}
{\bf 4. CONCLUDING REMARKS}%This is optional; you may or may not have this section.
\end{center}

In this paper, we consider a wave tank that has a wavemaker on one
side and an absorbing beach on the other side. A linear theory for
the flap type of wavemaker based on the full equation for linear
water waves is presented. It has been discussed for both the
single-flap and the double-flap wavemaker motion. By solving the
governing equation with its corresponding boundary conditions, it
turns out that the surface wave elevation contains a progressive
wave part and an evanescent modes part, where the latter part
vanishes far from the wavemaker. Furthermore, there is an explicit
relation between the wave height of the generated surface wave
elevation in the far field and the stroke of the wavemaker. This
relation depends explicitly on the water depth, the hinge
position, and the wave frequency.

Therefore, we are able to solve both the direct problem, where the
wavemaker stroke is prescribed and the wave height is the outcome,
as well as the inverse problem, where the wave height is given and
the wavemaker stroke follows as a result. The latter case is more
likely to be the desired situation for performing experiments in
the laboratory, allowing one to control the wavemaker motion in
such a way that the desired wave field is realized.

We also apply linear theory to the case when the wavemaker has
double flaps. This type of the wavemaker is useful in generating
waves in a wider range of frequencies. The double-flap wavemaker
has two actuators that move the upper flap and the main flap.
Normally, the upper flap is used to generate the wave components
with a frequency higher than a certain critical frequency
$\omega^{\ast}$, while the main flap is used to generate the wave
components with frequencies below $\omega^{\ast}$. By applying the
Figure of Merit (FoM) on the wavemaker flaps, one can determine
the critical frequency $\omega^{\ast}$, see equation
(\ref{critical_quantities}). The FoM depends only on the wave tank
depth and the position of the hinges. For a certain wave tank
configuration, the critical frequency has to be determined. Hence,
one can decide when to use either the upper flap or the main flap.
Since the calculation is based on linear theory, linear
superposition can be used to construct an irregular wave field.

This wavemaker theory can be extended using nonlinear theory. This
will involve mode generation, as a consequence of nonlinear
interaction among monochromatic wave components at a higher order.
This nonlinear extension is expected to be executed in similar
research in the future.

\vspace{1.5cc}
\noindent{\bf Acknowledgement.} %Acknowledgements may be presented here, if any.
This work has been executed at the Industrial \& Applied
Mathematics R \& D Group or Kelompok Penelitian dan Pengembangan
Matematika Industri dan Terapan Institut Teknologi Bandung (KPP
MIT - ITB), Indonesia and is supported by the Small Project
Facility of the European Union Jakarta, entitled `Building
Academia-Industry Partnership in the Sectors of Marine and
Telecommunication Technology'. The authors want to thank Dr.
Andonowati and Prof. E. van Groesen for inviting them to
participate in this project. Part of this work has been executed
at the University of Twente, the Netherlands, under the project
`Prediction and generation of deterministic extreme waves in
hydrodynamic laboratories' (TWI.5374) of the Netherlands
Organization of Scientific Research NWO, subdivision Applied
Sciences STW. The helpful discussions throughout the execution of
this research with Dr. Ren\'e Huijsmans are very much appreciated.
The second author also wishes to thank MARIN, the Netherlands,
for their support in the above mentioned STW project.

%Below is the list of references that are cited in the text. Do not include any
%paper that are not cited in the text. Make sure that the details are correct.

\vspace{2cc}
\begin{center}
{\small\bf REFERENCES}
\end{center}

\newcounter{ref}
\begin{list}{\small \arabic{ref}.}{\usecounter{ref} \leftmargin 4mm
\itemsep -1mm}

%Fill the list of references in the alphabetical order here. You may use
%\thebibliography instead, and/or labels. Adopt the following style:

%for article in a journal:
\bibitem{Dalzell} {\small {\sc J. F. Dalzell}, ``Analysis of Articulated Flap
Wavemakers'', {\it Technical Report TR75 - 1, ABA Electromechanical
Systems}, Pinellas Park, Florida, USA, March 1975.}

\bibitem{Dean} {\small {\sc R. G. Dean and R. A. Dalrymple}, {\it Water Wave Mechanics
for Engineers and Scientists}, volume \textbf{2} of Advanced
Series on Ocean Engineering, World Scientific, Singapore, 1991.}

\bibitem{Galvin} {\small {\sc C. J. Galvin, Jr}, ``Wave Height Prediction for Wave Generators
in Shallow-Water'', {\it Tech Memo {\bf 4}, U.S. Army, Coastal
Engineering Research center}, March 1964.}

\bibitem{Gilbert} {\small {\sc G. Gilbert, D. M. Thompson \& A. J. Brewer},``Design
Curves for Regular and Random Wave Generators'', {\it J. Hydraulic
Res.} {\bf 2}, no. 2, 163 -- 196, 1971.}

\bibitem{Press} {\small {\sc W. H. Press, B. P. Flannery, S. A. Teukolsky}, {\it Numerical Recipes
in FORTRAN: The Art of Scientific Computing}, Cambridge University
Press, Cambridge, 1992.}
\end{list}

\vspace{1cc}

%Type author(s)' name, institution and e-mail address(es) here.
%For example:

{\small\noindent {\sd W.M. Kusumawinahyu}: Department of
Mathematics, Institut Teknologi
Bandung, Jl. Ganesha 10 Bandung 40132, Indonesia.\\
Department of Mathematics, Universitas Brawijaya, Jl. May. Jend.
Haryono 169 Malang 65145,
Indonesia.\\
\vspace{0.75cc}
\noindent \hspace*{-0.25cm} E-mail: muharini@dns.math.itb.ac.id\\
{\sd N. Karjanto}: Applied Analysis and Mathematical Physics
Group, Department of Applied Mathematics, University of Twente,
P.O. Box 217, 7500 AE, Enschede, The Netherlands.\\
\vspace{0.75cc}
\noindent \hspace*{-0.25cm} E-mail: n.karjanto@math.utwente.nl\\
{\sd G. Klopman}: Applied Analysis and Mathematical Physics Group,
Department of Applied Mathematics, University of Twente, P.O. Box 217, 7500 AE, Enschede, The Netherlands.\\
Albatros Flow Research, Voorsterweg 28, 8316 PT, Marknesse, The Netherlands.\\
\noindent E-mail: g.klopman@math.utwente.nl}

\end{document}